# Media Usage Survey:
## Overall Comparison of Faculty and Students

Gerd Gidion, Michael Grosch
Institute for Vocational Education and General Education
Department of Vocational Education and Training
Karlsruhe Institute of Technology
Karlsruhe, Germany

Luiz Fernando Capretz, Ken Meadows
Department of Electrical and Computer Engineering
Teaching Support Centre
Western University
London, Canada

*Abstract*—Recent developments in the use of technologies in education have provided unique opportunities for teaching and learning. This paper describes the results of a survey conducted at Western University (Canada) in 2013, regarding the use of media by students and instructors. The results of this study support the assumption that the media usage of students and instructors include a mixture of traditional and new media. The main traditional media continue to be important, and some new media have emerged as seemingly on equal footing or even more important than the traditional forms of media. Some new media that have recently been in the public spotlight do not seem to be as important as expected. These new media may still be emerging but it is not possible to know their ultimate importance at this point. There was some variation in media usage across different Faculties but perhaps not as much variation as might have been expected.

I. INTRODUCTION

The integration of IT media and services in higher education has led to substantial changes in the ways in which both students and instructors study, learn, and teach [1], [2], [3]. Accordingly, a survey of students' and instructors' media usage habits has been conducted at Western University in 2013. This survey purports to measure the extent to which media services are used in teaching and learning as well as to assess changes in media usage patterns. The survey is a landmark, as it is the first of its kind in Canada and represents the initial foray into the North American post-secondary sector.

For the purposes of this research, media is defined as technology that supports and extends human communications. The conveyance of information represents a unidirectional form of communication and, therefore, information services is included in this definition. In the field of digital media, where the content is not attached to a physical data carrier, media services include both software and hardware. Because software media can be transferred to different hardware, the latter is necessary for software access, and thus, hardware constitutes an integral component of this definition.

The survey focuses primarily on the media usage habits of students and instructors. Based on assessing the way in which media use relates to academic education, that means teaching and studying. The identification of trends aims to provide an evidence base upon which more reliable predictions can anticipate future trends of media usage in higher education. The basic idea is that current academic education is utilizing (and influenced by) media, that are a combination of traditional (e.g., printed books and journals) and new (e.g., Google and Wikipedia) media. The actual situation has developed from former media usage habits, and these habits might change with the introduction of new media.

Partial results involving instructors and students only in the Faculty of Engineering were presented at the Canadian Engineering Education Association [4]. Other more focused survey on mobile learning maturity and specific for m-learning have been carried out recently [5], [6], [7]. Short-term academic education will likely be influenced by the level of satisfaction of media usage habits [8], [9]. This media usage survey was created to provide educational researchers with a deeper and more detailed understanding of students' and instructors´ technology usage in learning and of possible environmental factors that may influence that usage. This survey intended to incorporate the entire spectrum of media services, focusing on the following objectives:

- Evaluating media use in detail, including media use frequency, satisfaction with, and acceptance of both internal or university-provided and external services, print media, electronic text, social media, information technology, communication media, e-learning services, and IT hardware.
- Determining factors that might influence media use in learning, such as cultural differences, age, sex, and academic level as well as identifying similarities among student media usage.
- Creating a knowledge base for universities to understand the media usage of students and instructors as well as establishing a longitudinal international survey on technology use in tertiary education.
- Assessing prospective media trends and supporting the definition of media development as one of the strategic ideas at universities.
- Evaluating user satisfaction; thus media quality is also evaluated by measuring the acceptance of services used by students and instructors.

## II. RESEARCH METHODOLOGY

The survey comprises a fully standardized anonymous questionnaire containing a total of 150 items. Specifically, the tool measures usage frequency and user satisfaction with 53 media services, including:

- Media hardware and web connection, such as Wi-Fi, notebooks, tablet computers, desktop computers, and smartphones.
- Information services, such as Google search, Google Books, library catalogues, printed books, e-books, printed journals, e-journals, Wikipedia, open educational resources, and bibliographic software.
- Communication services, such as internal and external e-mail, Twitter, and Facebook.
- e-learning services and applications, such as learning platforms and wikis.

The survey comprises a fully standardized anonymous questionnaire containing a total of 150 items. Specifically, the tool measures usage frequency and user satisfaction with 53 media services, including:

These variables, as well as the previously mentioned methodology, were also used to create acceptance value. Additional variables underwent evaluation, such as some aspects of learning behavior, media usage in leisure time, educational biography, and socio-demographic factors.

The survey tool was first developed in 2009 and used at Karlsruhe Institute of Technology (KIT) in Germany [10]. During the application of the 15 follow-up surveys that were administered internationally, the original survey underwent optimization, translation into several languages, and validation. In this study, the survey was administered at Western University to undergraduate students and faculty members in January and February of 2013. The instructor survey, which resembles the student questionnaire, intends to compare the media usage of students and instructors by examining possible divergences in media culture that may create problems in the use of media for studying and teaching.

Initial invitations to participate in the research and two reminders were sent by email. Both faculty and student surveys were voluntary and anonymous, as indicated in the cover letters. For the student survey, three emails were sent by the Office of the Registrar staff to a stratified random sample of undergraduate and graduate students enrolled on the main campus in the Winter 2013 academic term. The faculty survey involved a similar procedure and targeted faculty teaching on the main campus during the Winter 2013 academic term. The data for this survey was collected online using an established survey provider: Unipark.

In the period between January 16th and February 15th 2013, 19978 students were invited to respond to the survey. Subsequently, exactly 1584 visits occurred on the survey website. Among the invited students, 1266 started to answer the questions, 985 completed the survey, and 803 recorded a completion rate of more than 90%.

In the period between January 29th and February 28th 2013, approximately 1400 instructors were solicited by email to answer the survey. During this time, exactly 332 visits occurred at the survey website. Although 252 faculty members started to answer the questions, 210 of them completed the survey. While participants were randomly selected from a broad spectrum of demographic characteristics and faculties, female students were more heavily represented in terms of respondents. Otherwise, with some caveats, respondents are generally regarded as representative of the January and February 2013 student and instructor population at Western. A summary of participation is shown in Table 1.

TABLE 1. Response numbers for Western's students and instructors that answered the question regarding the faculty of their primary area of study or primary teaching assignment.

| Faculty/ School | Students U/G | | | | Instructors | | | |
|---|---|---|---|---|---|---|---|---|
| | Population | | Participants | | Population | | Participants | |
| | N | % | n | % of 792 | N | % | n | % of 187 |
| Arts and Humanities | 1,232 | 5.7 | 82 | 10.4 | 151 | 11.0 | 15 | 8.0 |
| Education | - | - | - | - | 37 | 2.7 | 4 | 2.1 |
| Engineering | 1,310 | 6.0 | 56 | 7.1 | 94 | 6.8 | 11 | 5.9 |
| Health Sciences | 3,246 | 15.0 | 125 | 15.8 | 133 | 9.7 | 21 | 11.2 |
| Information and Media Studies | 969 | 4.5 | 45 | 5.7 | 44 | 3.2 | 3 | 1.6 |
| Law | - | - | - | - | 33 | 2.4 | 2 | 1.1 |
| Music | 527 | 2.4 | 37 | 4.7 | 44 | 3.2 | 15 | 8.0 |
| School of Business | 1,097 | 5.1 | 15 | 1.9 | 111 | 8.1 | 11 | 5.9 |
| School of Grad/Postdoc Studies | - | - | - | - | - | - | 2 | 1.1 |
| School of Med.&Dent. | 2,425 | 11.2 | 19 | 2.4 | 281 | 20.5 | 42 | 22.5 |
| Science | 4,244 | 19.6 | 173 | 21.8 | 203 | 14.8 | 23 | 12.3 |
| Social Science | 6,627 | 30.6 | 237 | 29.9 | 241 | 17.6 | 38 | 20.3 |
| Missing (this item) | | | 193 | | | | 23 | |
| Total | 21,677 | | **985** | | 1372 | | **210** | |

## III FINDINGS

A survey was developed and deployed to collect data, which was analyzed quantitatively, presented below. Of particular interest to a primary co-investigator, a software engineer, was that students from Engineering were only significantly different than their fellow students in their frequency of usage of a small number of media (e.g., computer games). Instructors show some differences in their reported media usage, but there were notable similarities as well, such as the seemingly pervasive use of Google search.

*A. Some General Media, Learning, and Studying Habits*

In the comparison between students and instructors some general media, learning and studying / teaching habits showed up to be different. Attending class is slightly more relevant for students, instructors seem to work more frequently with computers than students use their computer to study. Instructors utilize the internet as intensive as students and more. Students visit libraries more than instructors (with the exception of the beginners and students from Engineering). Instructors work more with printed material they found themselves than students and work together with colleagues more often than students study with other students.

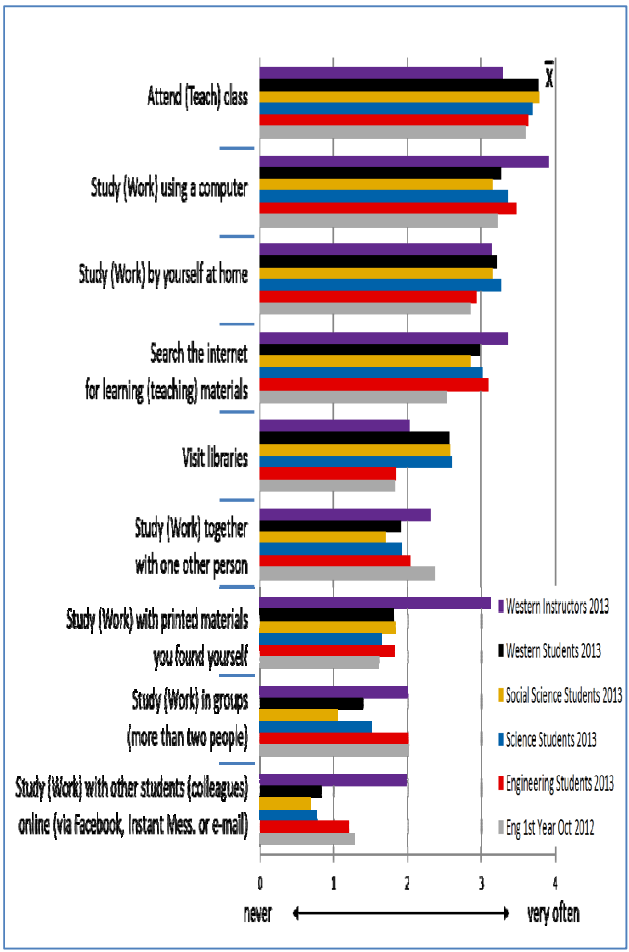

Fig. 1. Means of students´ and instructors´ responses to the question: How often do you do the following? (different version of the question for instructors in brackets), the question was rated on a five-point Likert scale with the following choices: 'never' (0), 'rarely' (1), 'sometimes' (2), 'often' (3), and 'very often' (4); the figure shows the means of all instructors (valid n = 210) and all students (valid n = 985) and of 5 subgroups, related to their faculty; general participation: Social Science: n = 237; Science: n = 173; Engineering: n = 56; Engineering first year students October 2012: n = 100)

*B. Media Usage in Free Time*

Concerning the usage frequency of media in free time the results show that students use Facebook and video sharing websites more often, whilst instructors read more books. Students play computer games more and instructors seem to work – on a lower level – with Google+ more than students do.

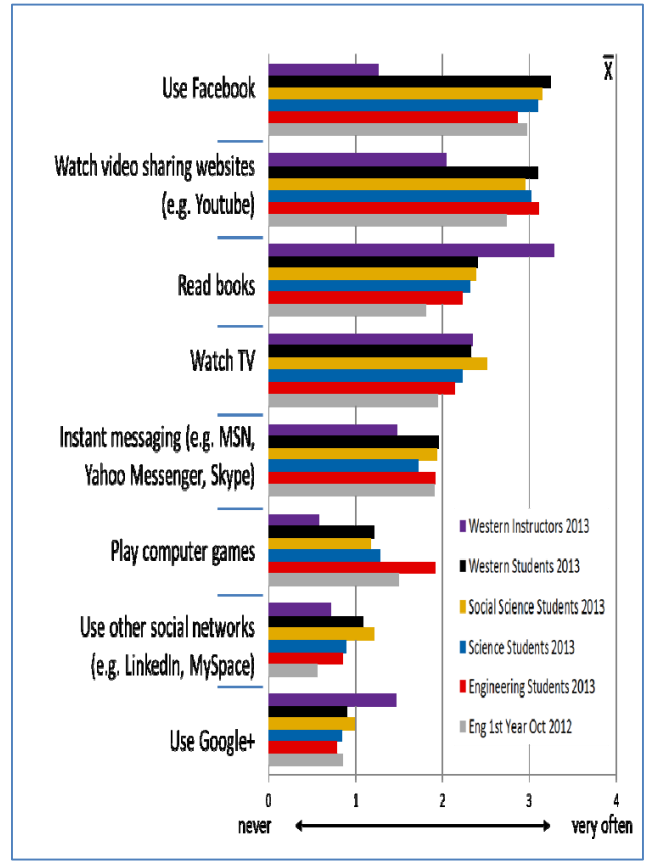

Fig. 2: Means of students´ and instructors´ responses to the question: How often do you do the following during your free time? (different version of the question for instructors in brackets), the question was rated on a five-point Likert scale with the following choices: 'never' (0), 'rarely' (1), 'sometimes' (2), 'often' (3), and 'very often' (4); the figure shows the means of all instructors (valid n = 210) and all students (valid n = 985) and of 5 subgroups, related to their faculty; general participation: Social Science: n = 237; Science: n = 173; Engineering: n = 56; Engineering first year students October 2012: n = 100)

*C. Frequency of e-Learning Application Usage*

Looking at the items concerning e-Learning applications the students results show higher values as instructors for online (self) tests and online exams, instructors results came to a little higher value in the items 'e-Learning applications as part of a course' and 'learning / educational software'.

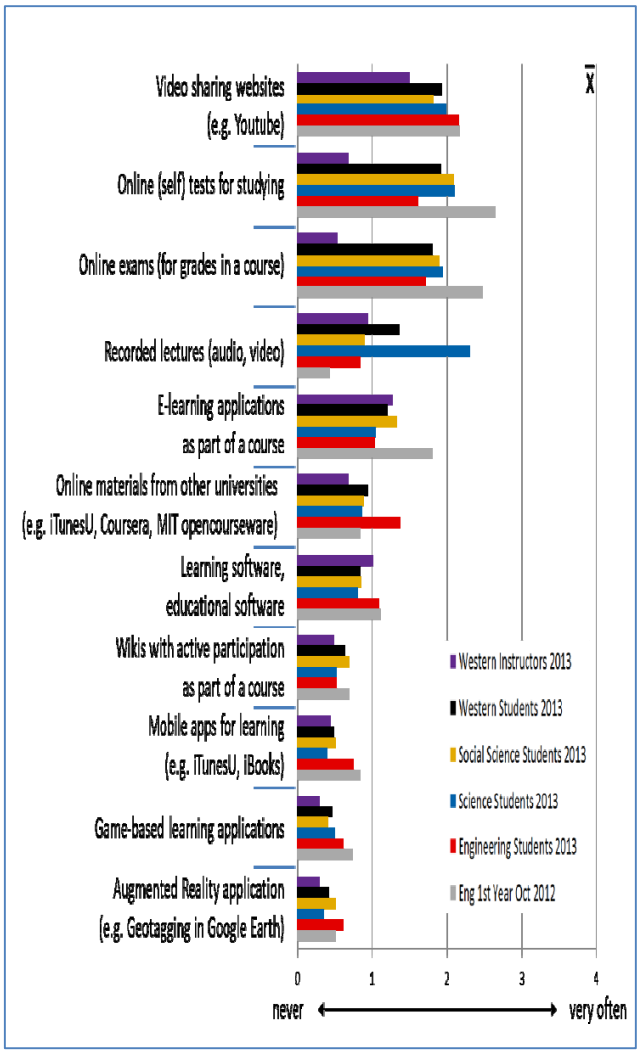

Fig. 3. Means of students´ and instructors´ responses to the question: How often do you use the following for learning - studying / your academic work (i.e. teaching, research, service)? (different version of the question for instructors in brackets), the question was rated on a five-point Likert scale with the following choices: 'never' (0), 'rarely' (1), 'sometimes' (2), 'often' (3), and 'very often' (4); the figure shows the means of all instructors (valid n = 210) and all students (valid n = 985) and of 5 subgroups, related to their faculty; general participation: Social Science: n = 237; Science: n = 173; Engineering: n = 56; Engineering first year students October 2012: n = 100)

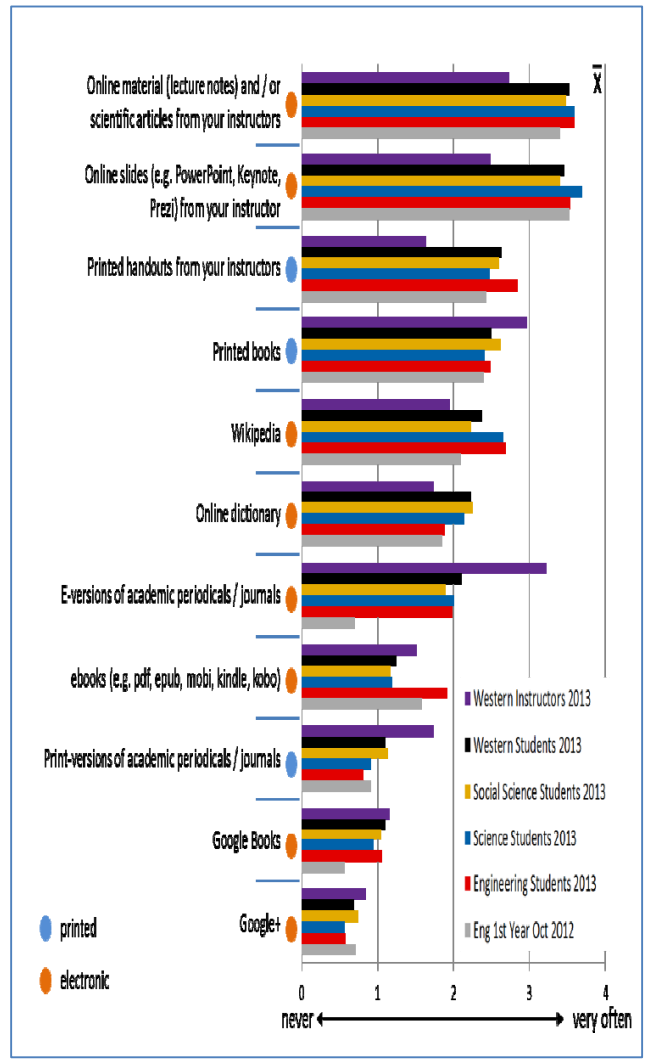

Fig. 4. Means of students´ and instructors´ responses to the question: How often do you use the following for learning - studying / your academic work (i.e. teaching, research, service)? (different version of the question for instructors in brackets), the question was rated on a five-point Likert scale with the following choices: 'never' (0), 'rarely' (1), 'sometimes' (2), 'often' (3), and 'very often' (4); the figure shows the means of all instructors (valid n = 210) and all students (valid n = 985) and of 5 subgroups, related to their faculty; general participation: Social Science: n = 237; Science: n = 173; Engineering: n = 56; Engineering first year students October 2012: n = 100)

### D. Frequency of Printed vs Electronic Media

In the group of items concerning printed vs electronic media usage the online and the printed documents from instructors are more often used by students, the same as Wikipedia and online dictionaries. Instructors came to higher usage values for printed books, academic journals, ebooks, Google books and Google+.

### E. Frequency of Usage of Social Network Related Publications

The usage frequency of social network related applications is similar between students and instructors for Google search, somewhat lower for instructors regarding Wikipedia, Newsgroups / internet forums and social bookmarking, lower on students side for Google+ and other social networks like LinkedIn. Facebook and Twitter seem to be far more used by students than by instructors.

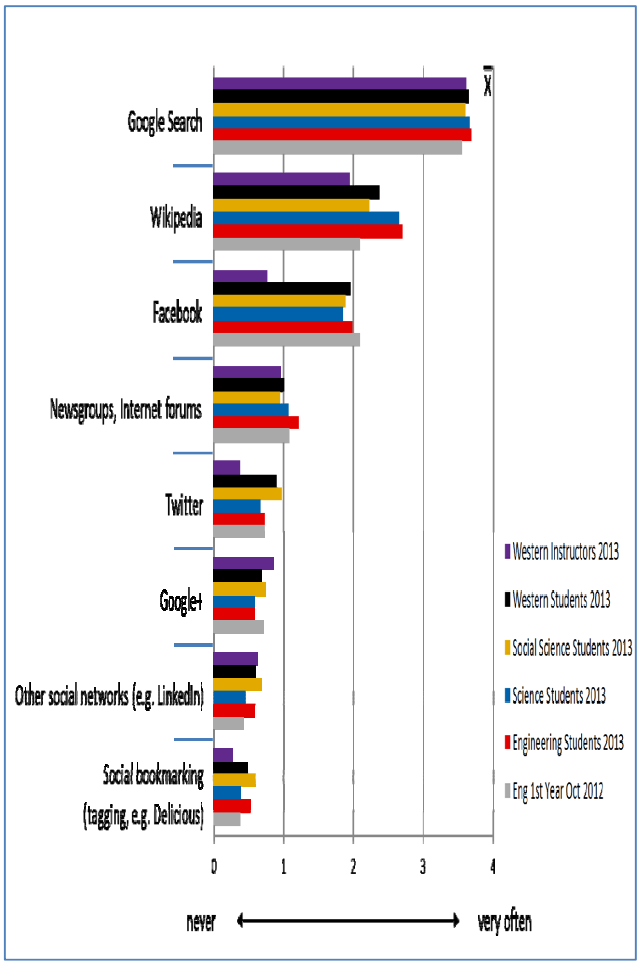

Fig. 5. Means of students´ and instructors´ responses to the question: How often do you use the following for learning - studying / your academic work (i.e., teaching, research, service)? (different version of the question for instructors in brackets), the question was rated on a five-point Likert scale with the following choices: 'never' (0), 'rarely' (1), 'sometimes' (2), 'often' (3), and 'very often' (4); the figure shows the means of all instructors (valid n = 210) and all students (valid n = 985) and of 5 subgroups, related to their faculty; general participation: Social Science: n = 237; Science: n = 173; Engineering: n = 56; Engineering first year students October 2012: n = 100)

### F. Satisfaction with the Usage of Social Network Related Applications

In the group of items concerning printed vs electronic media usage the online and the printed documents from instructors are more often used by students, the same as Wikipedia and online dictionaries. Instructors came to higher usage values for printed books, academic journals, ebooks, Google books and Google+.

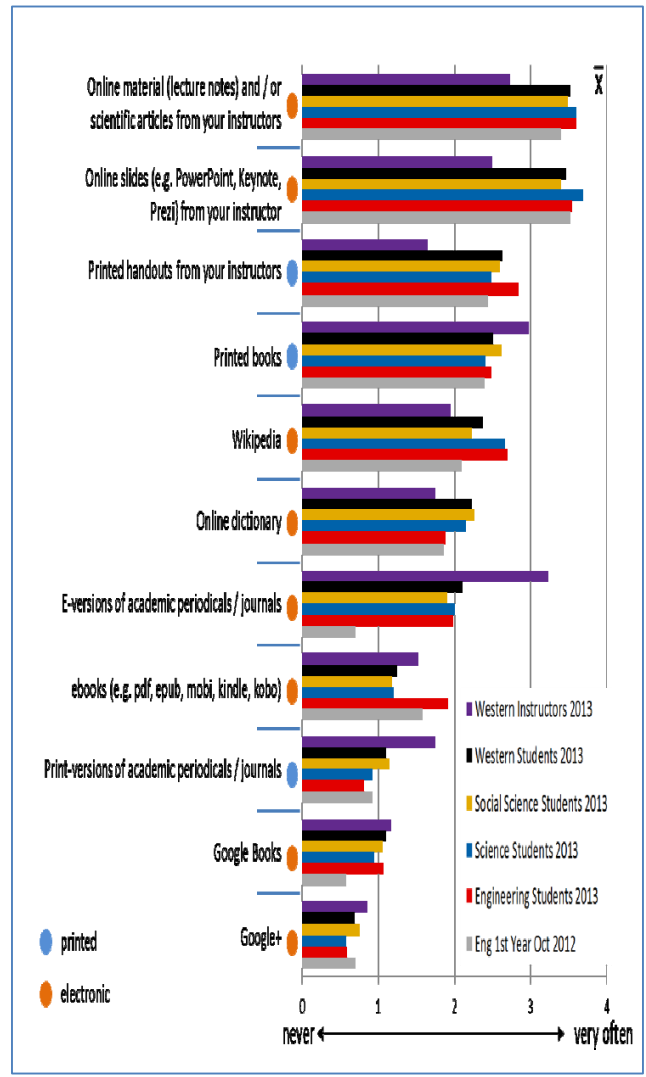

Fig. 6. Means of students´ and instructors´ responses to the question: How often do you use the following for learning - studying / your academic work (i.e., teaching, research, service)? (different version of the question for instructors in brackets), the question was rated on a five-point Likert scale with the following choices: 'never' (0), 'rarely' (1), 'sometimes' (2), 'often' (3), and 'very often' (4); the figure shows the means of all instructors (valid n = 210) and all students (valid n = 985) and of 5 subgroups, related to their faculty; general participation: Social Science: n = 237; Science: n = 173; Engineering: n = 56; Engineering first year students October 2012: n = 100)

### G. Satisfaction with the Usage of Printed vs Electronic Media

The comparison of satisfaction values for printed vs electronic shows all items with the exception of Google+ on high positive level, and a few differences between instructors and students like in the item 'e-versions of academic periodicals / journals', where not only instructors seem to be more satisfied than students, but the result from Engineering students shows an even higher value.

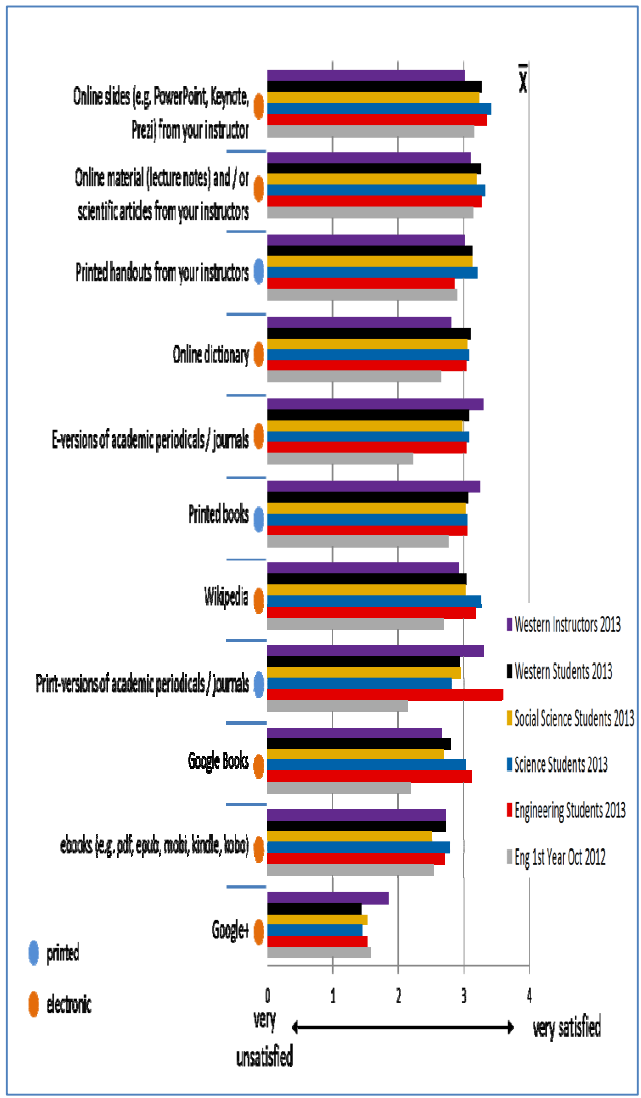
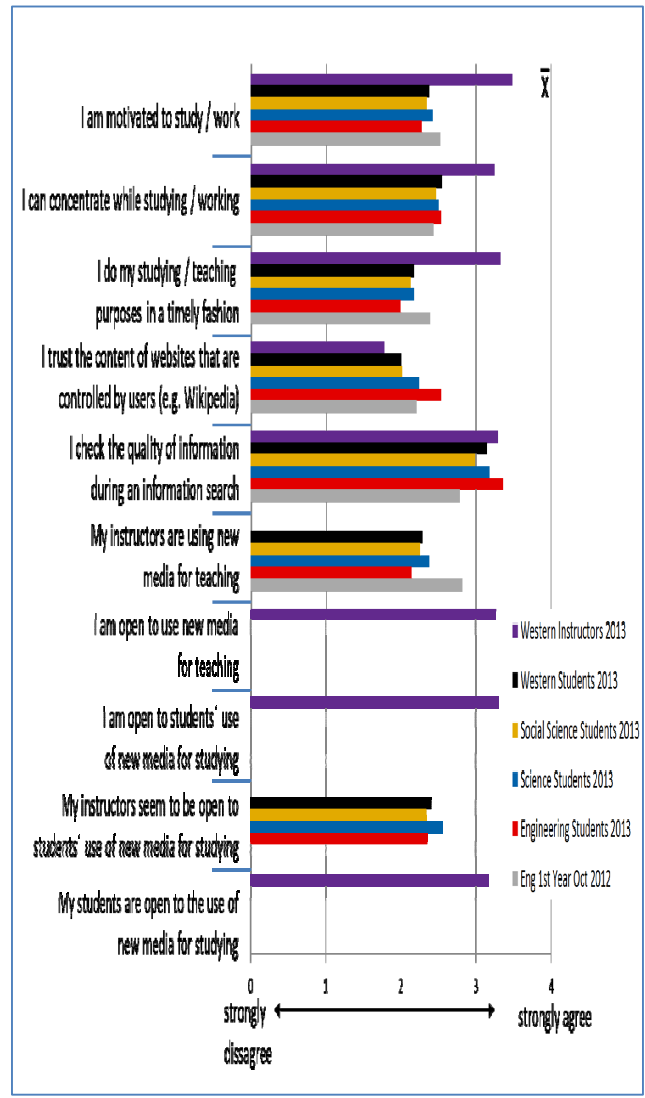

Fig. 7. Means of students´ and instructors´ responses to the question: If you use it: How satisfied are you with the use / functionality of the following for learning - studying / your academic work (i.e., teaching, research, service)? (different version of the question for instructors in brackets), the question was rated on a five-point Likert scale with the following choices: 'never' (0), 'rarely' (1), 'sometimes' (2), 'often' (3), and 'very often' (4); the figure shows the means of all instructors (valid n = 210) and all students (valid n = 985) and of 5 subgroups, related to their faculty; general participation: Social Science: n = 237; Science: n = 173; Engineering: n = 56; Engineering first year students October 2012: n = 100)

Fig. 8 Means of students´ and instructors´ responses to the question: To what extent do you agree /disagree with the following statements? (different version of the question for instructors in brackets), the question was rated on a five-point Likert scale with the following choices: 'never' (0), 'rarely' (1), 'sometimes' (2), 'often' (3), and 'very often' (4); the figure shows the means of all instructors (valid n = 210) and all students (valid n = 985) and of 5 subgroups, related to their faculty; general participation: Social Science: n = 237; Science: n = 173; Engineering: n = 56; Engineering first year students October 2012: n = 100)

### H. Some Attitudes Related to Media Usage

The group of items with statements about some attitudes related to media usage shows a heterogeneous picture. Instructors seem to be motivated to work notably more than students are to study, and they see their ability to concentrate on a higher positive level than students think about theirs. The same result can be stated for the item 'I do my studying (for instructors: teaching) purposes in a timely fashion'. Some items were just used in the instructors resp. in the students survey, e.g. the openness for the use of new media. Instructors see themselves generally as more open than students see them.

### IV CONCLUSIONS

Looking at the survey results, it can be stated that several traditional media are still very relevant and continuing to be in high use, however, in a changing environment. Printed material and slides from the instructors as well as printed books were deemed to have high values of usage frequency and satisfaction. Attending class and visiting libraries are often performed habits, and the university's own services are more frequently used than external academic sources.

At the same time, additional new media, such as the electronic versions of material from instructors or the Learning

1019

Management System, are established and utilized with a similar intensity. It seems that these newly established media, which are based on traditional media, are very easy and comfortable to access and use and, therefore, in the future they are likely to be used more often than their traditional counterparts.

This intensive use of new media services and arrangements might be a phenomenon enabled by new habits that encourage working with media. Students and instructors are equipped with mobile and continuously network connected computers, and they are proficient in using them from their experience (and self-organized learning) in their private life. The use of some media can be stated as obligatory, especially the use of Google Search; that is on the highest rank of frequency of use as well as one of the highest satisfaction values with the usage. Differences in usage exist between students and instructors and between free time and studying usage. The use of Facebook and YouTube shows very high values of usage frequency, so this might also be stated as a habit.

Certain innovative usage variations of new media for teaching and learning/studying are distinct – such as wikis as a part of a course, recorded lectures, or online tests – but more often for certain courses. They have been developed, launched, and proved; however, just a few arrangements seem to apply to these options. It can be assumed, that in those cases where a serious effort has been made, these new variations of working with new media have a distinct relevance, such as recorded lectures in science.

Media usage expands the interdependence with the market of academic education. So the competition with other universities and service providers has intensified. Although the frequency of use of online materials from other universities (e.g., iTunesU, Coursera, MIT open-courseware) or mobile apps for learning has not reached a similar level as Western's own materials, the use of media with a non-direct competitive influence seems to be especially remarkable, such as video sharing websites, Wikipedia or Google books. It can be assumed that the competition will be much more intense in the future, because the main players on the market collect (and utilize) much more specific data about students and instructors than every single university can (or would be allowed to) do.

Potentially arising future media and trends cannot be identified with this survey, but quite new media like Google+, augmented reality applications, or game-based learning applications might be more important, although not very common in use, for teaching and studying at the moment. In addition to that, side effects of some of the established and ubiquitous usage of some media will very probably lead to some consequences. So the habit of working with Google Search facilitates, so-called 'hyper targeting', and creating electronic user profiles that will perhaps be used for technology-based customization and delivery of services at a high level of situational individualization.

Overall, the media usage by students and instructors is in some aspects different, but explainable, too, as in the case of Desktop PCs, Facebook, and YouTube. Instructors – as a heterogeneous group – generally have a more traditionally oriented usage of media, but some show ingenuousness in using new options. So the frequency of using Google+ is higher for the instructors as compared to the students. Many new media are extensively used by both instructors and students and can be considered as 'new habits' (in a world of academia, where some habits seem to be unchangeable, although that has been intended over the years).

Students from different Faculties show a general similarity. Significant differences can be noted in the comparison between two different faculties, like Arts and Humanities vs. Science or Engineering (e.g., with the frequency of reading books of Arts and Humanities), but this seems to be explainable, too. Additionally gender has a significant influence, especially in the frequency of use of the so-called 'Social Media'.